\documentclass[a4paper,3p]{elsarticle}
\usepackage{amsmath}
\usepackage{bm}
\usepackage{cancel}

\usepackage{txfonts}
\usepackage[utf8x]{inputenc}
\usepackage[protrusion=true,expansion=true]{microtype}
\usepackage{hyperref,xcolor}
\usepackage[T1]{fontenc} 
\usepackage{graphicx}
\usepackage{tikz}
\usepackage{enumitem}
\usepackage{color} 
\usepackage[mathlines]{lineno} 
\usepackage{array,multirow,makecell}
\usepackage[normalem]{ulem}
\usepackage{ulem}

\usepackage{lineno}

\newcommand\simpst{\bgroup\markoverwith{\textcolor{magenta}{\rule[0.5ex]{2pt}{0.4pt}}}\ULon}

\def\rbm#1{\xrbm#1\relax^\relax\valign}
\def\xrbm#1^#2\relax#3\valign{%
\mathbf{#1}\ifx\valign#2\valign\else^{\mathbf{#2}}\fi}

\begin{document}
\begin{frontmatter}

\title{ Analytical planar wavefront reconstruction and error estimates
\\ for radio detection of extensive air showers}

\author[a,b]{Arsène Ferrière}
\ead{arsene.ferriere@cea.fr}
\author[c,d]{Simon Prunet}
\author[a,e]{Aurélien Benoit-Lévy}
\author[b,d]{Marion Guelfand}
\author[d,e,f]{Kumiko Kotera}
\author[g,h]{Matías Tueros}

\address[a]{Universit\'{e} Paris-Saclay, CEA, List, F-91120, Palaiseau, France}
\address[b]{Sorbonne Université, CNRS, Laboratoire de Physique Nucléaire et des Hautes Energies (LPNHE), 4 Pl. Jussieu, 75005 Paris, France}
\address[c]{Université Côte d’Azur, Observatoire de la Côte d’Azur, CNRS, Laboratoire Lagrange, \\ Bd de l’Observatoire, CS 34229, 06304 Nice cedex 4, France}
\address[d]{Sorbonne Universit\'{e}, UPMC Univ.  Paris 6 et CNRS, UMR 7095, Institut d'Astrophysique de Paris, 98 bis bd Arago, 75014 Paris, France}
\address[e]{Department of Physics, Pennsylvania
State University, University Park, PA, USA}
\address[f]{Vrije Universiteit Brussel, Physics Department, Pleinlaan 2, 1050 Brussels, Belgium}
\address[g]{IFLP - CCT La Plata - CONICET, Casilla de Correo 727 (1900) La Plata, Argentina}
\address[h]{Depto. de F\'isica, Fac. de Cs. Ex., Universidad Nacional de La Plata, Casilla de Coreo 67 (1900) La Plata, Argentina}

\begin{abstract}

When performing radio detection of ultra-high energy astroparticles, the planar wavefront model is often used as a first step to evaluate the arrival direction of primary particles. This model estimates the direction by adjusting the wavefront orientation based on the peak timing of the signal traces from individual antennas. However, despite it's simplicity, the usefulness of this approach is limited by the lack of a good assessment of its robustness and the lack of confidence in its performance.\\
To address these limitations, this study presents two analytical methods to solve for the planar wavefront arrival direction. In addition, we provide the corresponding analytical reconstruction uncertainty, offering a more a more detailed evaluation of the reconstruction's reliability.
\end{abstract}


\end{frontmatter}


\section{Introduction}
\label{introduction}

An extensive air shower (EAS) is produced when a high-energy particle (cosmic ray, gamma ray, neutrino or one of its secondary particles) interacts with molecules of the atmosphere, generating a cascade of daughter particles. In air, this development can be used to detect the highest energy astroparticles for instance via the direct collection of the daughter particles, or the shower radiation production in Cherenkov, fluorescence light, or radio. For all these techniques, determining the arrival direction of the primary particle is key. First because it serves as a fundamental discrimination against terrestrial emissions. Second because it can be a useful ingredient to reconstruct the other parameters of the air-shower, such as its energy or the nature of the primary particle. The arrival direction can also be used to discern cosmic rays from neutrinos, mostly emerging from below the horizon. Finally, this has become crucial in the current multi-messenger astronomy era, for source identification and follow-up searches.

In this work, we focus on the radio-detection technique. The electromagnetic component of the shower undergoes charge separation due to the geomagnetic field and other less prominent processes. The resulting varying current leads to an emission in the radio regime, which can be detected by simple antennas arrays (see e.g., reviews by \cite{Huege_2016,Schroder_2017}). This concept has been successfully tested with various experiments like LOPES \cite{Falcke_2005}, CODALEMA \cite{Ardouin_2005} or LOFAR \cite{van_Haarlem_2013}. 
These instruments have focused on {showers} arriving with low inclination with respect to the zenith. 
The arrival direction can be reconstructed by making use of the timing information collected at each antenna in the array, and fitting a wavefront model, accounting for the varying light propagation speed in the atmosphere. A robust determination of the initial radiation emission direction thus requires a correct modeling of the wavefront. 

The shape of the wavefront as observed on the ground depends strongly on the ratio between the emission region size and the distance to the detector. {Showers close to the zenith} present rather close-by emission-point configurations: geometrically, one can infer that the wavefront curvature is quasi-spherical close to the shower axis, and becomes conical further out, leading to a hyperbolic function. The pioneering LOPES \cite{Apel_2014} and CODALEMA \cite{Ardouin_2005} experiments found that plane or hyperbolic wave modelings enabled indeed reasonable reconstructions. The efficiency of the hyperbolic modeling was further confirmed by LOFAR analysis \cite{Corstanje_2015}. On the other hand, showers arriving with very inclined zenith angles, as targeted by next-generation radio detectors like GRAND \cite{GRAND_paper_2019} or BEACON \cite{beacon_2023}, imply that the observer is located at large distances from the emission region. Recent studies \cite{Decoene_2023} have demonstrated that the wavefront shape could then be modeled accurately by a spherical function. 

The planar wavefront (PWF) model {has been used extensively by various experiments, either as the main technique for direction of arrival reconstruction \cite{Ardouin_2005, Ardouin_2011, Charrier_2019}, or as an initial step to reduce the parameter space before considering more involved modeling such as a spherical wavefront model \cite{Decoene_2023}.}

We present in this work a full analytical methodology to calculate the arrival direction using the planar wavefront model, and assess its robustness and efficiency. Most importantly, we precisely evaluate the errors associated to this computation. Such estimates are lacking in the literature: the source positions are reconstructed without associated, theoretically-motivated uncertainties. On the contrary, this study provides the community with both reliable direction reconstruction and uncertainty estimations. The simplicity of the planar wavefront modeling allows for a fast and analytical computation, which could be implemented online at first detector triggering levels for event selection after precise timings have been determined by independent methods.
The source code implementing the methods described below is publicly available in \cite{Ferriere2024}.

The layout of this paper is as follows. 
In Section~\ref{section:methods} we present our methodology, its formalism, the solutions and the error estimation calculations. 
In Section~\ref{section:performances}, we evaluate the performances of our method, and discuss the outcomes in Section~\ref{section:conclusions}.

\section{Methodology}\label{section:methods}
\subsection{Problem formulation and solution}
\begin{figure}
    \centering
    \includegraphics[scale=0.47,clip]{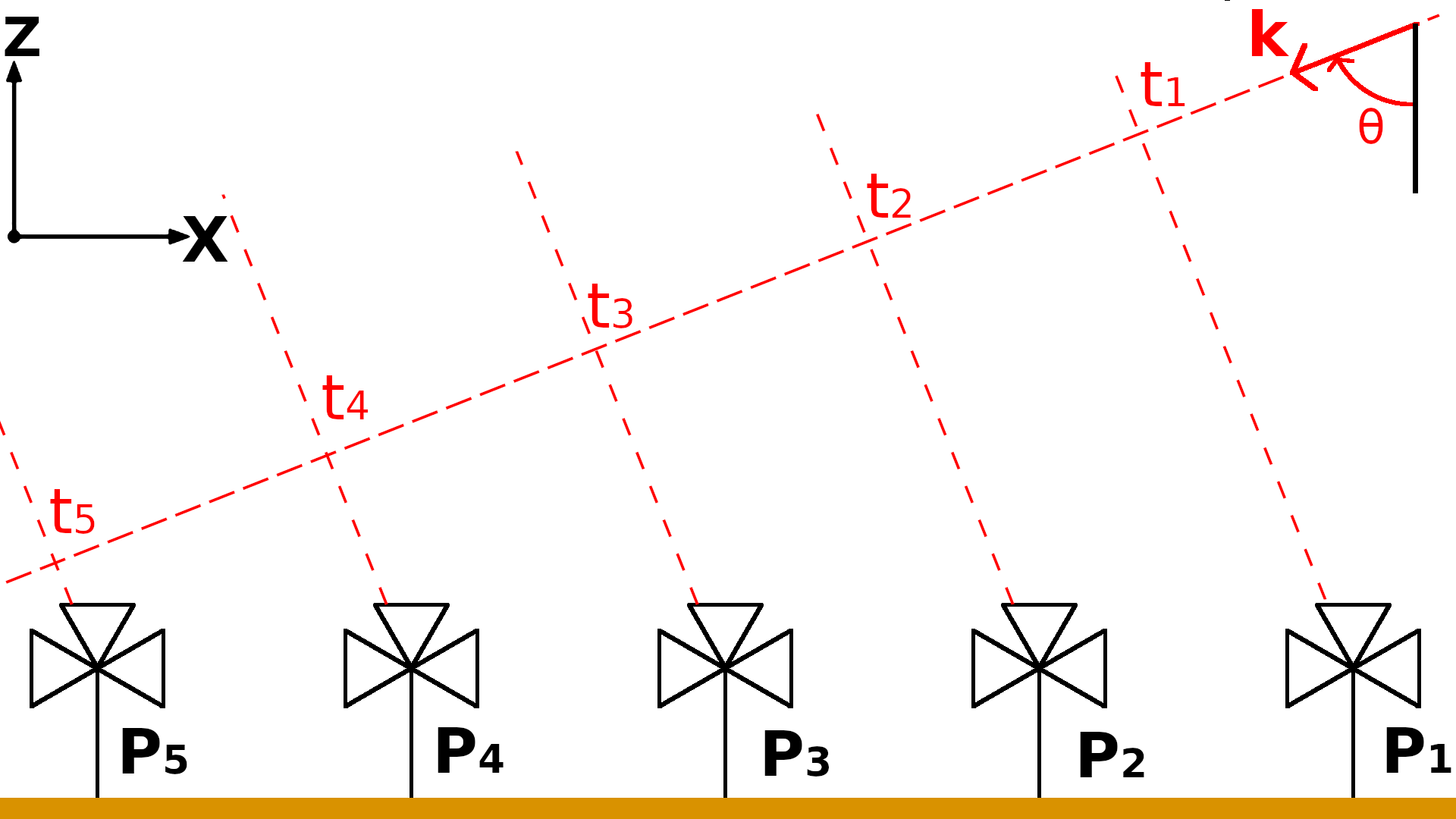}
    \caption{2d representation of the planar wavefront model. The cosmic ray (coming from the right),
generates a wave with a wavefront perpendicular to the direction of propagation. All antennas on a same
perpendicular plane to the direction of propagation will be triggered at the same time.
   }
    \label{fig:pwf}
\end{figure}
Under the assumption that the wavefront is planar, as depicted on Fig.~\ref{fig:pwf}, the time of arrival on antenna $i$ can be related to the direction vector $\rbm k $, the position of the antenna $\rbm{p}_i$, and  a time of reference $t_0$ as follows:
\begin{equation}\label{eq:PWF_model}
    t_i = \frac{1}{c'} \rbm p_i \cdot \rbm k  + t_0 \text{;  } \ \rbm k = 
        \left(\begin{matrix}
            k_x\\
            k_y\\
            k_z
\end{matrix}\right)\ ,
\end{equation}
where $c'$ is the average speed of light in the traversed medium.
\setlength{\fboxsep}{0pt}%
\begin{figure}
    \centering
    \fbox{\includegraphics[width=.5\linewidth, trim={5cm 0 1cm 6.2cm},clip]{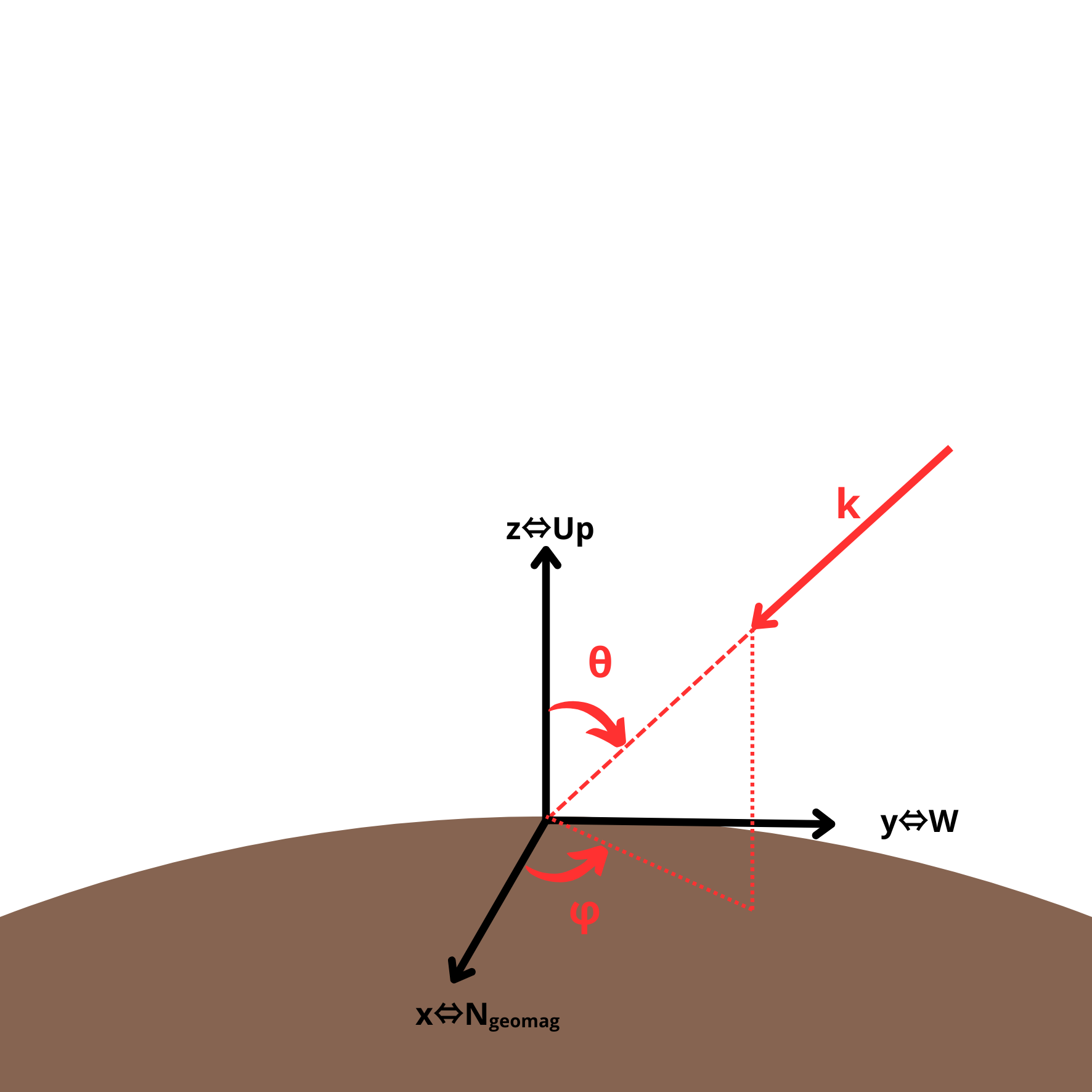}}
    \caption{{Angle convention used in this study. The fixed reference frame is oriented along the axis
down up, south north, east west. The angle $\theta$ and $\phi$ for zenith and azimuth are the polar coordinate
angles of the vector indicating the direction from where the particle is coming from. $\rbm{k}$ is in the opposite direction i.e. the direction of propagation of the shower.}}

    \label{fig:angle_convention}
\end{figure}
Here $\rbm k$ is a unit vector as the signal must propagate at the speed of light. It is perpendicular to the wavefront and in the direction of propagation of the signal. It directs the shower axis in the PWF case. With the angle convention
used (see Fig.~\ref{fig:angle_convention}), we define the zenith angle $\theta$ and azimuth angle $\phi$
of the air shower by the following equation:
\begin{equation}
\rbm k = 
\left(\begin{matrix}
            k_x\\
            k_y\\
            k_z
\end{matrix}\right)\ =
\left(\begin{matrix}
            - \sin(\theta)\cos(\phi)\\
            - \sin(\theta)\sin(\phi)\\
            - \cos(\theta)
\end{matrix}\right)\ .
\end{equation}

{In presence of timing uncertainties}, Equation~\ref{eq:PWF_model} can be written in vector format to account for $n$ antennas:
\begin{equation}\label{eq:PWF_model2}
    \rbm{T} = \rbm {P}\, \rbm k + \rbm{\epsilon}\ ,
\end{equation}
where $\rbm{\epsilon}$ is the noise impinging each antenna, {considered a centered Gaussian of covariance matrix $\rbm Q$}: $\rbm{\epsilon} \thicksim \mathcal{N}\left(\rbm 0,\ \rbm Q \right)$\footnote{$\mathcal{N}(\mu, \Sigma)$ denotes the Gaussian distribution with mean $\mu$ and covariance matrix $\Sigma$.},
 
$\rbm T$ is a length n vector encapsulating all timings : $\rbm{T} = c' [t_i-\bar{t}]_{i=\{1..n\}}$, 
and $\rbm{P} = [\rbm p_i-\bar{\rbm p}]_{i=\{1..n\}}$ is a $n \times 3$ matrix. $\Bar{t}$ and $\Bar{ \rbm p}$ are the weighed averages of the times and positions of the antennas:
{
 \begin{eqnarray}
    \bar{\rbm p} &=& \frac{\rbm{1_N}^T\rbm{Q}^{-1}[\rbm p_i]_{i=\{1..n\}}}{\rbm{1_N}^T \rbm Q^{-1} \rbm{1_N}}, \nonumber \\ 
    \text{and}\label{zero_mean}&&\\
    \bar{t} &=& \frac{\rbm{1_N}^T\rbm{Q}^{-1}[t_i]_{i=\{1..n\}}}{\rbm{1_N}^T \rbm Q^{-1} \rbm{1_N}}. \nonumber
\end{eqnarray}
with $\rbm{1_N}$ is a column vector of 1's. 
If we consider uncorrelated noise, the noise covariance matrix is diagonal: 
\begin{equation}
    \rbm Q = \text{diag}(\sigma_1^2, \sigma_2^2, ..., \sigma_n^2)
\end{equation}
with $\sigma_i$ the standard deviation of the timing noise on antenna $i$. The weighted means become:
\begin{eqnarray}
    \bar{\rbm p} &=& \frac{1}{\sum_i{\sigma_i}^{-2}}\sum_i\frac{\rbm p_i}{{\sigma_i}^2} \nonumber \\ 
    &&\text{and}\\
    \bar{t} &=& \frac{1}{\sum_i{\sigma_i}^{-2}}\sum_i\frac{t_i}{{\sigma_i}^2} \nonumber
\end{eqnarray}

}
{Operations performed in Eqs.~\ref{zero_mean} essentially zero-mean the antennas and timings vectors $\rbm T$ and $\rbm P$ and remove the dependence on the time of reference $t_0$ that we do not need to estimate.}

{From Eq.~\ref{eq:PWF_model2}, $\rbm T$ follows  a Gaussian distribution, with mean $\rbm{Pk}$ and variance $\rbm Q$. The likelihood function then reads:}
\begin{equation}\label{eq:likelihood}
    \rm p(\rbm T\, |\, \rbm P,\rbm k) = \frac{\exp{ \left[-
     \frac{1}{2}  (\rbm T-\rbm P \,\rbm k)^T \rbm Q^{-1} (\rbm T-\rbm P\, \rbm k) \right]}}{{\left({2\pi \det{\rbm Q}}\right)}^{n/2}}\ .
\end{equation}

In order to estimate the direction of arrival $\rbm k$, we need to find the maximum of the likelihood function, or equivalently, find the minimum of the negative log-likelihood function:

{\begin{align}
L_{ \rbm P, \rbm T}(k) &= -\log \rm p(\rbm T|\rbm P,\rbm k )  \nonumber \\ 
 &=\frac{1}{2}  (\rbm T-\rbm P\rbm k )^T  \rbm Q^{-1}   (\rbm T-\rbm P\rbm k ) + A  \nonumber \\ 
 &=\frac{1}{2}  \left(\rbm{k}^T\rbm \Sigma^{-1}\rbm{k} -2\rbm{b}^T\rbm{k}\right) + A' 
\label{eq:Cost}
\end{align}
under the constraint that 
\begin{equation}
    \Vert \rbm k \Vert = 1
\end{equation}
Where $A$ and $A'$ are constants that do not interfere in the minimization process and with:
        \begin{equation}
            \begin{cases}
            \rbm \Sigma = \left(\rbm P^T \rbm Q^{-1} \rbm P\right)^{-1} \\
            \rbm b = \rbm P^T \rbm Q^{-1} \rbm T. \\
            \end{cases}
        \end{equation}}
The solution to this process will be called $\rbm{k_{MLE}}$.

We present two methods to solve this optimization problem. The first one, a projection method, {is most useful when the height variations of the antenna positions are small compared to the horizontal spread of their positions. This is the case for most surface radio detectors.} The second method is a semi-analytical method and offers a more general solution. 
    
\subsubsection{Projection method} 
The minimum of the function in Eq.~\ref{eq:Cost} without the constraint that $\Vert \rbm k  \Vert = 1$ is given by:
    \begin{equation}
        \rbm k ^{*} = (\rbm P^T\rbm Q^{-1}\rbm P)^{-1}\rbm P^T\rbm Q^{-1} \rbm T {= \rbm \Sigma \rbm b}\ \ ,
    \end{equation}
    as it is a linear regression problem.

\begin{figure}
        \centering
        \includegraphics[scale=0.3,clip]{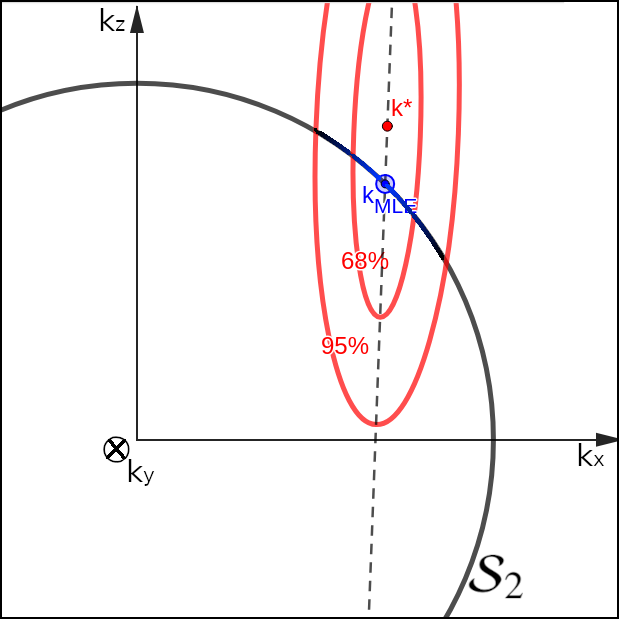}
        \caption{Visualization of a cross section of the regression problem. The unconstrained linear regression produces $\rbm k^*$ and the distribution represented by the red contour lines. The projection approximation projects $\rbm k^*$ on the sphere along the dashed line (the large axis of the distribution). It is an approximation of $\rbm{k_{MLE}}$, the maximum of likelihood on the unit sphere $\mathcal{S}_2$. The likelihood distribution under constraint is represented by the blue color gradient on the sphere. The distributions represented here are not to scale.
        }
        \label{fig:visu_distrib}
    \end{figure}

    Since the timing measurement errors are Gaussian, we can also express the distribution of $\rbm k $ {given noisy measurements $\rbm T$} by:
    \begin{equation}\label{eq:DistNoConstraint}
        \rbm k  \thicksim \mathcal{N}\left( \rbm{k^{*}},\ \rbm{\Sigma} \right)\,.
    \end{equation}
    This distribution is represented by the red contour lines in Figure~\ref{fig:visu_distrib}. 
    It is skewed along an axis close to the vertical line (dashed line on Figure~\ref{fig:visu_distrib}) because the antennas lie almost on an horizontal plane. The component of $\rbm k^*$ {along this line} is reconstructed with low precision thus its covariance term is orders of magnitude higher than for the other components.
    
    To apply the constraint, we disregard this component and set its value so that $\Vert \rbm k \Vert=1$. On Fig.~\ref{fig:visu_distrib}, it amounts to projecting $\rbm{k^*}$ along the dashed line onto the unit sphere $\mathcal{S_2}$. More formally, as $\rbm \Sigma^{-1}$ is a positive-definite matrix,  it can be diagonalized into an orthogonal basis: $\rbm \Sigma^{-1} = \rbm{\Phi}\rbm{\Lambda}\rbm{\Phi}^T$ with $\rbm{\Lambda}$ a diagonal matrix composed of the eigen values of $\rbm \Sigma^{-1}$ in decreasing order and $\rbm{\Phi}$ an orthogonal transition matrix.
    {The density function} of the distribution given by Eq.~\ref{eq:DistNoConstraint} can then be written: 
    \begin{align}
        \rm p(\rbm k ) &= 
        C\exp\left[{-\frac{1}{2}\left(\rbm{\Phi}^T\rbm k - \rbm{\Phi}^T\rbm k^*\right)^T \rbm{\Lambda}\, \left(\rbm{\Phi}^T\rbm k -\rbm{\Phi}^T\rbm k^*\right)}\right] \\ 
        & \text{ with: } \bm \Lambda=
                    \left(\begin{matrix}
                        \lambda_1 & 0 & 0\\
                        0 & \lambda_2 & 0\\
                        0 & 0 & \lambda_3
                    \end{matrix}\right) \nonumber
    \end{align}
     The eigen values $\lambda_1$, $\lambda_2$, $\lambda_3$ are such that: $\lambda_3 \ll \lambda_1, \lambda_2$.

    Projecting $\rbm{\Phi}^T\rbm {k^*}$ onto the unit sphere and along the eigen vector associated with $\lambda_3$ gives the final prediction in the eigen basis coordinates system. Projecting the resulting vector in the original $(\rbm x,\rbm y,\rbm z)$ coordinate system by multiplying by $\rbm{\Phi}$ produces $\rbm{k_{proj}}$ which is an approximation of the maximum likelihood estimator. 
    
    If $\rbm{\Phi}^T\rbm k^*=\left[k_{x'}, k_{y'}, k_{z'}\right]^T$ then
    \begin{equation}
        \rbm{k_{proj}} = \rbm{\Phi}\left[k_{x'},\ k_{y'},\ -\sqrt{1-(k_{x'}^2+k_{y'}^2)}\right]^T     \ .
    \end{equation}
    In the case where $k_{x'}^2+k_{y'}^2 > 1$, we set $k_{z'}=0$ then normalize $\rbm{\Phi}^T\rbm k^*$ so that $k_{x'}^2+k_{y'}^2 = 1$. The negative sign before the square root favors negative values for $k_z$ as we expect down going EAS.    
    
    \subsubsection{Semi analytical method} 
        Eq.~\ref{eq:Cost} is a quadratic form that is convex in $\rbm{k}$ but the constraint that $\rbm{k}$ lies on the unit sphere makes the optimization problem non-convex.
        {
        However, the unit sphere is compact and the quadratic form is continuous, so we know that the optimization problem has a global minimum. }
        It can be shown \cite{Hager_2001} that $\rbm{k_{MLE}}$ is a solution of Eq.~\ref{eq:Cost} if and only if there exist $\mu$ such that $\rbm \Sigma^{-1} +\mu\mathbf{I}$ is positive semidefinite, and $(\rbm \Sigma^{-1} + \mu\mathbf{I})\rbm{k_{MLE}} = \rbm b$.
        
        Let us diagonalize the matrix, $\rbm \Sigma^{-1} = \mathbf{\Phi\Lambda\Phi}^T$ again, where $\rbm \Lambda$ is the diagonal matrix whose diagonal are the eigen values $[\lambda_1, \lambda_2, \lambda_3]$ of $\rbm \Sigma^{-1}$ in decreasing order, and define $\bm \beta = \rbm{\Phi}^T\rbm{b}$.
        
        The solution to Eq.~\ref{eq:Cost} is given by $\rbm{k_{MLE}} = \rbm{\Phi}\rbm{c}$
        where $\rbm{c}$ is computed in the following way:\newline let us define $\mathcal{E}_3 = \{i \mid \lambda_i= \lambda_3 \}$ and $\mathcal{E}_+ = \{ i \mid \lambda_i>\lambda_3\}$. We then have two distinct cases:
        
        \begin{itemize}
            \item{\emph{Degenerate case}.}
        If $\beta_i=0$ for all $i \in \mathcal{E}_3$, and 
        \begin{equation}
            \sum_{\mathcal{E}_+} \frac{\beta_i^2}{(\lambda_i - \lambda_3)^2} \leq 1,
        \end{equation}
        then $\mu = -\lambda_3$ and $c_i = \beta_i/(\lambda_i - \lambda_3)$ for all $i \in \mathcal{E}_+$. The values of $c_i$ for $i \in \mathcal{E}_3$ are chosen arbitrarily, as long as they ensure that $\rbm{k}$ is of unit norm: $\sum_{\mathcal{E}_3} c_i^2 = 1 - \sum_{\mathcal{E}_+} c_i^2$. 
        
        \item{\emph{Non degenerate case}.}
        In this case, we have $c_i = {\beta_i}/{(\lambda_i + \mu)}$, such that $\sum c_i(\mu)^2 =1$. The function on the left hand side of the last equation is strictly convex in $\mu$, in the interval $[ -\lambda_3 + \sqrt{\sum_{\mathcal{E}_3} \beta_i^2}, -\lambda_3 + \|b\| ]$ . $\mu$ can then be estimated with a line search algorithm.
        \end{itemize}

        Once $\rbm c$ is obtained, we can multiply by $\rbm \Phi$ to get $\rbm{k_{MLE}}$.
        The degenerate case corresponds to when the antennas lie on a common plane, the solution is the same as the approximate solution from the previous subsection. 

        For all previous methods, a final step is added to ensure the reconstructed $\rbm k$ points downward. When the predicted vector points upward, we reflect it through the plane formed by the antennas, whose normal vector is the eigen vector associated with $\lambda_3$.
        
\subsection{Uncertainty}\label{subsec:uncertainty_calcul}

    The classical linear regression with no constraint produces the distribution of $\rbm k$ given $\rbm P$ and $\rbm T$ : $\rm p(\rbm k|\,\rbm P, \rbm T)$ as shown in Eq.~\ref{eq:DistNoConstraint}.

    Applying the constraint amounts to looking for:
    \begin{equation}
        \rm p(\rbm k|\,\rbm P, \rbm T, \Vert \rbm k \Vert = 1)
    \end{equation} rather than $\rm p(\rbm k|\,\rbm P, \rbm T)$.
    As we show in Sect.~\ref{section:performances}, the reconstruction errors of the estimators proposed in the previous section are small (typically less than a degree). We can assume that the unit sphere is locally flat and approximated by its tangent plane at $\rbm{k_{MLE}}$. {Doing so, we can obtain an analytical expression for the covariance matrix of the reconstructed zenith and azimuth angles $\theta$ and $\phi$:     \begin{equation}
    \label{eq:variance}
        \Bar{\rbm \Sigma} = 
        \left[  
            \rbm{R_a}^T \rbm \Sigma^{-1} \rbm{R_a}
        \right]^{-1}
    \end{equation}
        with
        \begin{equation}
        \rbm{R_a} = \left(\begin{matrix}
            -\cos(\theta)\cos(\phi)& \sin(\theta)\sin(\phi) \\
            -\cos(\theta)\sin(\phi)& -\sin(\theta)\cos(\phi) \\
                       \sin(\theta)   &              0        \\
        \end{matrix} \right).
    \end{equation}
    The derivation for this covariance matrix is available in \ref{app:Cov_Matrix}.}
    In practice, we approximate $\rbm{R_a}$ by replacing $\theta$ by $\theta_{\rm{MLE}}$ and $\phi$ by $\phi_{\rm{MLE}}$, the zenith and azimuth angles of $\rbm{k_{MLE}}$ the maximum likelihood estimation obtain from any planar wavefront reconstruction method. 
    
    If we denote by $\psi$ the angle between the prediction $\rbm{k_{MLE}}$ and the target direction $\rbm k$:
    \begin{equation}
        \psi = \cos^{-1}(\rbm k\,\cdot\,\rbm{k_{MLE}} ).
        \label{eq:mean_error}
    \end{equation}
    We can estimate the mean value of $\psi^2$ as
    \begin{equation}
        \Vert \rbm{k_{MLE}} - \rbm k \Vert^2 = \psi^2 + o(\psi^3)
    \end{equation}
    thus 
        \begin{equation}
        <\psi^2> \approx <\Vert \rbm{k_{MLE}} - \rbm k \Vert^2> \approx \rm tr (\rbm C \Bar{\rbm \Sigma} \rbm C)
    \end{equation}
    with 
    \begin{equation*}
        \rbm C = \left(\begin{matrix}
            1 & 0 \\
            0& \sin(\theta_{\rm MLE})
        \end{matrix} \right).
    \end{equation*}
    This correspond to the uncertainty described in \cite{Ardouin_2011}:
    \begin{equation}
        <\psi^2> = \sigma_{\theta\theta}^2 + \sin^2{(\theta)}\,\sigma_{\phi\phi}^2
    \end{equation}
    with { $\sigma_{\theta\theta}^2$ and $\sigma_{\phi\phi}^2$}, the diagonal values of $\Bar{\rbm \Sigma}$, which correspond to the variance of the estimators of $\theta$ and $\phi$. 
\section{Performances}\label{section:performances}

To evaluate the performances of the direction estimators as well as the uncertainty estimator, we will use two distinct sets of simulations.
    \begin{figure}
        \centering
        \includegraphics[scale=0.5,clip]{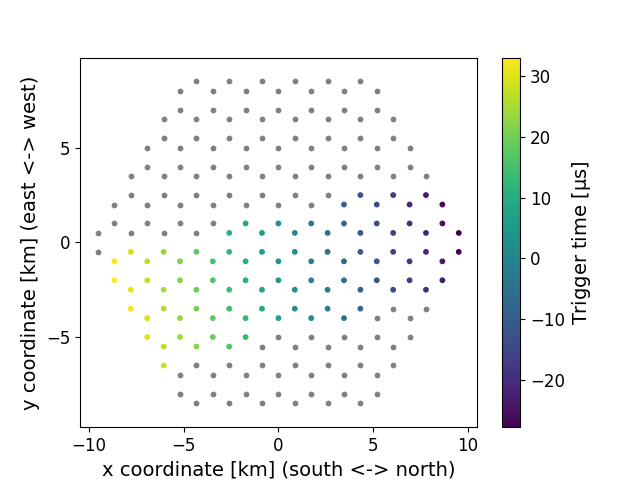}
        \caption{Antenna layout used in the study. Antennas are placed on an hexagonal grid with 1000\ m side length. Every event only triggers a subset of all the available antennas. Here, the layout intersected a cosmic ray with a zenith angle $\theta=87.1$°, azimuth $\phi=14.5$° and energy $E=4.0$ EeV. The triggered antennas are color-coded according to the arrival times.
        }\label{fig:AntennaLayout}
    \end{figure}
\begin{figure*}[t]
    \centering
    \includegraphics[scale=0.5]{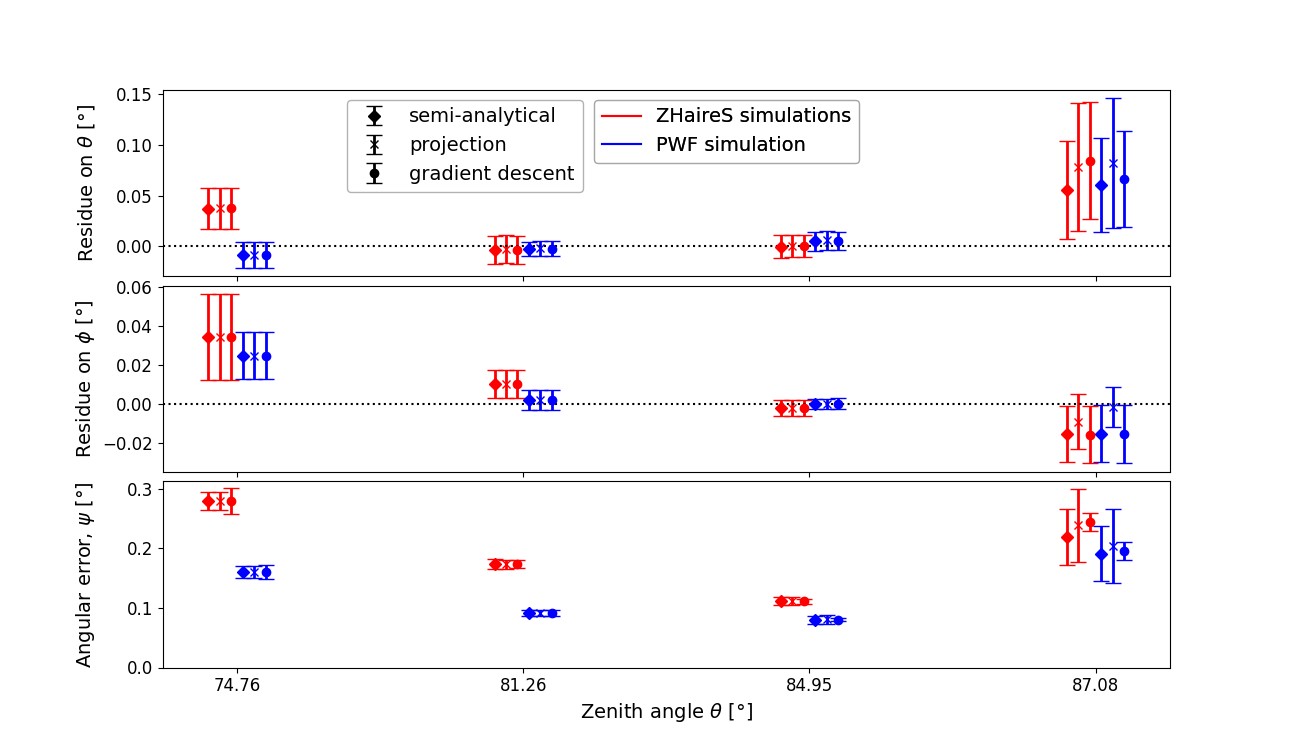}
    \caption{Mean zenith (top) and azimuth (middle) angle reconstruction residues on the PWF and ZHAireS simulation sets. The bottom plot represents the mean angular error, defined in Eq.~\ref{eq:mean_error} The error bars represent one standard deviation of the mean value estimator.}
    \label{fig:perfPWF_v2}
\end{figure*}
    \begin{itemize}
    \item We first use a set of {555} ZHAireS simulations \cite{Alvarez_Mu_iz_2012}. This set of simulations comes from a larger set of 1500 events generated with the same simulation procedure as the one described in \cite{layout2024}. These are simulations of the electric field sampled every 0.5ns at selected antenna locations (see Fig.~\ref{fig:AntennaLayout}). {These antennas are placed on an grid of hexagons with sides of 1 km. The grid is 20 km wide and presents a total height variation $\Delta z = 260$ m,  much smaller than the horizontal span $\Delta x = 20$ km}. All the events have been generated with a discrete number of input parameters:
    \begin{itemize}
        \item 5 values of zenith angle $\theta$ : [63.01°, 74.76°, 81.26°, 84.95°, 87.08°].
        \item 5 values of energy $E$ : [0.1, 0.3, 0.63, 1.6, 4.0] EeV
        \item 3 types of primary type : [proton, gamma, iron]
    \end{itemize} 
    {For each triplet of values for $\theta$, $E$ and primary type, 20 simulations with random azimuth angle $\phi$ between 0° and 180° have been generated. }
    To make data closer to antenna measurements, we do a series of transformations to the data. First, we filter the traces to keep only frequencies between 50\,MHz and 200\,MHz with a band-pass filter. We extract the maximum amplitude and time of maximum of the Hilbert envelope of the resulting traces. Antennas with signal amplitude lower than $90\, \rm{\mu}$V/m are disregarded as it would not trigger the trace recording, according to typical noise levels. {This cut is above 4 times the standard deviation of the amplitude of the background noise, considered at $22\, \rm{\mu}$V/m. It would ensure very few false triggering on real data. }Furthermore, any EAS that triggered less than 5 antennas according to previous trigger criterion are rejected. After these quality cuts, we are left with the 555 simulated events. {No events with 63.01° zenith angle have passed the quality cuts.}

    {Gaussian noise with a standard deviation $\sigma = 10$ ns  is added to extracted timings of every antenna to represent the noise introduced by the GPS and measurements uncertainties. The noise is independent and identically distributed for all antennas for testing purposes but it can be set differently to account for correlations between antennas and for different noise amplitudes.}
    This new set of simulation let us assess the performances on data as close as possible to what a radio antenna would measure from a EAS. 

    This dataset will be referred as the ZHAireS simulation set in the following sections.
    \item
    We consider a second set of simulations based on the planar wavefront modeling. From {all the 555} event of the ZHAireS simulation set, we create a corresponding one with the same triggered antennas. The sole difference is that we replace the wavefront arrival times by times generated by a planar wavefront with the same zenith and azimuth angles. Similarly as the ZHAireS set, we add a Gaussian noise with a standard deviation of 10\ ns.     
    This second simulation set is useful to show the performance of the methods in an idealized situation where the only errors come from the stochasticity of the data and the solver approximations but not from a discrepancy between the simulation and the model used to reconstruct the arrival direction. 
    In addition, it let us generate many realizations of the same event but with different noise values. We will refer to this set of simulations as the PWF simulation set.
    \end{itemize}
All simulations used to assess the performances of the methods have a zenith angle above 74°, which is very inclined compared to what has been considered in the past (\cite{Apel_2014}, \cite{Ardouin_2005}). We have focused on these highly inclined events because next-generation radio arrays are designed to be more sensitive at high zenith angles. However, the methods described are not restricted to this range of zenith angles.

\subsection{Estimator precision}

In this section, we compare the two methods defined in section~\ref{section:methods} with the more established and traditionally used gradient descent type techniques that are usually considered in this type of analysis \cite{Decoene_2023, Ardouin_2011}.

For each simulation set, we apply the projection, semi-analytical and gradient descent estimators to the simulated arrival times, and present the binned residues in the top and middle panels of Fig.~\ref{fig:perfPWF_v2}. 
For a proper reconstruction we use a fixed refractive index $n_{\rm{atm}}=1.00014$ (see ~\ref{app:IRrefinement} for more details).

As can be seen, the two new methods presented in this work give similar and consistent results (diamond and cross symbols) compared to the gradient descent (dot symbol).
It should be noted that all the methods also provide residues consistent with zero, meaning that they do not exhibit significant bias on both simulation sets. While this is expected for the PWF simulation set, which is generated using the same model as the estimators, the fact that this is also the case for the ZHAireS set demonstrates the PWF formalism, despite its simplicity is a very good descriptions of very inclined EAS.

On the bottom panel of Fig.~\ref{fig:perfPWF_v2}, we present the mean angular errors for the three estimators and two simulation sets. This is the mean value of the angle $\psi$ as described in section~\ref{section:methods}.
This error is always positive and describe the pointing precision of the estimator.
On average, the angular error is around 0.2° for all zenith angles.

\subsection{Uncertainty calibration}

\begin{figure}
    \centering
    \includegraphics[scale=0.5]{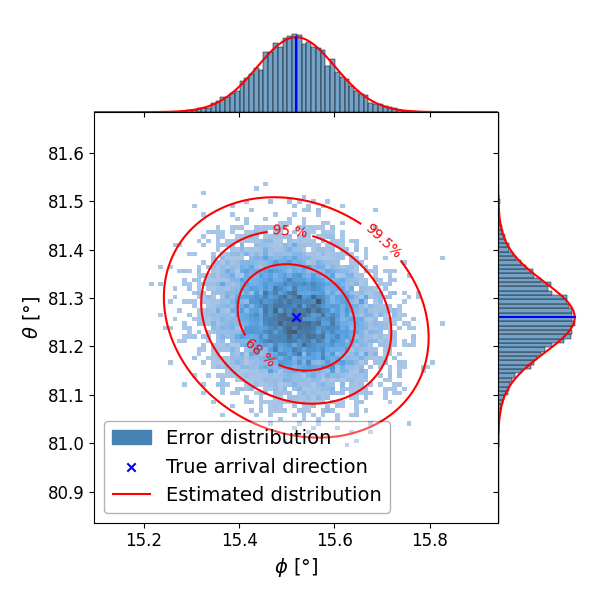}
    \caption{Example of confidence regions prediction, the uncertainty estimation closely matches the error distribution. The blue distribution is the histogram of many reconstruction of the same event ($\theta = 81.26$°, $\phi=15.52$°) with different noise realizations. The red lines are the isolines of the estimated uncertainties.}
    \label{fig:confidence_regions}
\end{figure}

In this section, we investigate the calibration of the theoretical uncertainties that we computed in Sect.~\ref{section:methods} and verify their calibration. Essentially, uncertainties are well calibrated when the reported variance, e.g. if we assume a Gaussian distribution, matches the empiric variance, which is determined by considering the dispersion of the results of the estimator.
We illustrate this on Fig.~\ref{fig:confidence_regions} where we compare the theoretical uncertainties from Eq~\ref{eq:variance} to the dispersion of reconstructed events for the PWF simulation set.
More specifically, we considered one specific event from the PWF simulation  set (this event has $\theta = 81.26$°, and $\phi=7.69$°), and generated 10000 different timing noise realizations, resulting in 10000 different events coming form the same direction. We applied the semi-analytical estimator to each of these events, and show the 2-d histogram of the reconstructed events in blue on Fig.~\ref{fig:confidence_regions}. We also show the 1-d histogram along each angle coordinate. We show in red the 68\%, 95\%, and 99.5\% theoretical contours, derived from Eq.~\ref{eq:variance}, as well as the 1-d probability function on top of the 1-d blue histogram. As can be seen, both distributions are consistent.

\begin{figure}
\centering
\includegraphics[scale=0.5,clip]{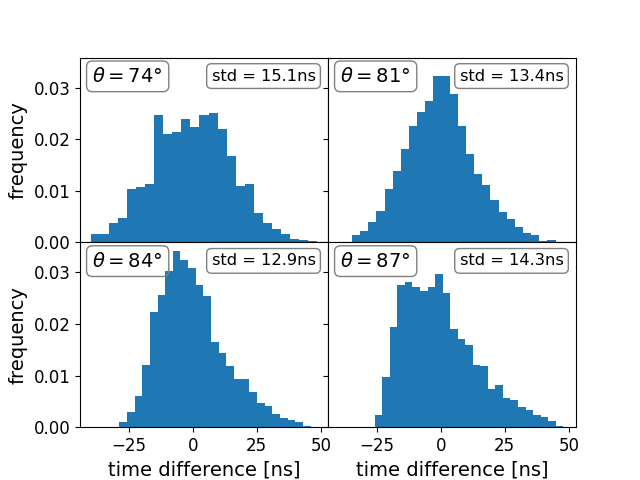}
\caption{Histogram of the difference of arrival times between the ZHAireS and PWF simulation set for the four zenith bins that are considered.
}
\label{fig:schema_example_deltat}
\end{figure}

We now turn our attention to the calibration on the ZHAireS simulation set. On Fig.~\ref{fig:schema_example_deltat} we show, for each zenith bin, the histogram of the differences between the arrival times in the raw ZHAireS simulations (before filtering and noise) and the one we would obtain with a planar wavefront. These histograms illustrate the fact that the wavefront simulated by ZHAireS isn't perfectly planar, they represent the deviation of the simulated wavefront from an idealized planar wavefront.
For each zenith bin, we add the variance of these time differences, denoted $\sigma_{\rm model}^2$, to the Gaussian timing noise variance, resulting in adding an effective systematic Gaussian noise to the previous noise distribution:
\begin{equation}
\sigma_{\rm{tot}}^2 = \sigma_{\rm model}^2 + \sigma^2 \ .
\end{equation}

\begin{figure}
\centering
\includegraphics[scale=0.5,clip]{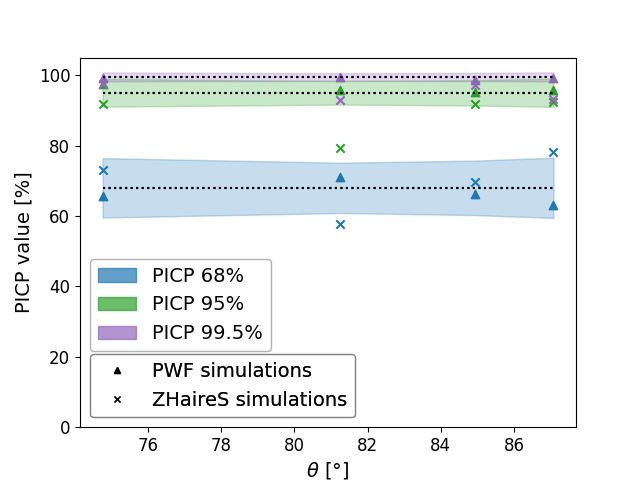}
\caption{Prediction Interval Coverage Probability (PICP) for 68\%, 95\% and 99.5\% quantiles on PWF dataset (triangle) and ZHAireS dataset (cross). The uncertainty estimator is well calibrated on PWF dataset but under-confident around 82° on the ZHAireS simulations. The shaded region are 95\% confidence intervals of the PICP estimator (2 standard deviations). } 
\label{fig:JoinedPICP}
\end{figure}

More generally, we can measure the calibration of uncertainties with a prediction interval coverage probability (PICP) diagram. For a chosen percentage, say $68\%$, we measure on our dataset the proportion of events where the true arrival direction lie inside the $68\%$ confidence region given by the theoretical uncertainties. If those are correctly calibrated, we expect this proportion to be close to $68\%$. 
We represent this diagram on Fig.~\ref{fig:JoinedPICP} for the PWF (triangles) and ZHAireS (cross) simulation sets for 68\%, 95\% and 99.5\% intervals. The envelop represents the statistical deviation tolerated (2 standard deviation) inside which we can consider that we are well calibrated. If the value is outside this envelop, we can confidently say the estimator isn't perfectly calibrated.
It is the case for zenith around $82$° {for ZHAireS simulations} but the PICP metric lies most of the time in the acceptable range around its target value.

\subsection{Computational efficiency}

Finally, we compare the computation efficiency of the different methods. The PWF reconstruction method described in \cite{Ardouin_2011} and \cite{decoene_sources_2020} use an alternative loss function:
\begin{eqnarray}
    L'_{\rbm P, \rbm T}(k) &=& \sum_{i,j}{ \left((t_i-t_j) - \frac{1}{c'}(\rbm p_i - \rbm p_j) \cdot \rbm k \right)^2 } \nonumber \\ 
    &=&  
    \Vert \rbm{\Delta T} - \rbm{\Delta P}\rbm{k} \Vert ^2
\end{eqnarray}
This loss function is mathematically equivalent to Eq.~\ref{eq:Cost} but has a time complexity of $\mathcal{O}( n^2 )$ while Eq.~\ref{eq:Cost} is in $\mathcal{O}(n)$ making it computationally faster. The minimization of this loss function with gradient descent with respect to $\theta$ and $\phi$ is calling the function $d$ times so the total complexity of commonly used method is $\mathcal{O}(dn^2)$. Conversely, the time complexity of the analytical-solutions or projection approximation is the time complexity of the linear regression or $\mathcal{O}(3^2 n + 3^3) = \mathcal{O}(n)$ (3 because we work in 3 dimension). The time complexity of inverting $\rbm Q$ is linear as $\rbm Q$ is a diagonal matrix. Furthermore, the complexity of finding $\mu$ in the semi-analytical method has very small impact on the performances.

The analytical estimators have much less computation to do, especially for large footprints and could therefore be used in an online configuration for direct reconstruction. A simple Python implementation on a personal computer gives timings of the order of 50 $\mu$s, which is subdominant for a typical trigger rate  of 1kHz \cite{Decoene_2019}.

\section{Conclusions, discussion}\label{section:conclusions}
We have proposed a mathematical and probabilistic approach to the planar wavefront description of air showers.
By construction, the direction of propagation of the wavefront is linearly dependent on the time of arrival and the positions of the antennas. We have demonstrated that despite the constraint that the propagation vector must be unitary, an analytical solution exists. A geometrical approach allows us to estimate the uncertainty on the reconstruction from the uncertainty on the measurements. The analytical uncertainty estimation for reconstruction also allows for the prediction of error bars and confidence regions for direction estimators that are reliable for a wide range of zenith angles. This reconstruction method can be fast and robust and used as a proxy for more complex estimators as well as an online reconstruction method for triggering procedure or first direction estimation.

\section*{Acknowledgements}

We thank the GRAND Collaboration for fruitful discussions, and in particular Valentin Decoene and Anastasiia Mikhno. We are also grateful to Valentin Niess for his review of this paper and helpful comments. This work was supported by the French Agence Nationale de la Recherche (PIA ANR-20-IDEES-0002; France), the CNRS Programme Blanc MITI ("GRAND" 2023.1 268448; France), CNRS Programme AMORCE ("GRAND" 258540; France), the Programme National des Hautes Energies of CNRS/INSU with INP and IN2P3 co-funded by CEA and CNES. KK acknowledges support from the Fulbright-France program. Simulations were performed using the computing resources at the CCIN2P3 Computing Centre (Lyon/Villeurbanne – France), partnership between CNRS/IN2P3 and CEA/DSM/Irfu. 

\appendix
\bibliographystyle{elsarticle-num}

\bibliography{biblio}

\appendix
\section{Uncertainty matrix}\label{app:Cov_Matrix}

In this section, we derive the covariance matrix of our uncertainties on the values of $\theta$ and $\phi$ the zenith and azimuth angle of the primary. Let's assume we have an estimation of $\rbm{k_{MLE}}$, the maximum likelihood estimator of the propagation vector $\rbm k$. We also assume we know that the distribution of the noise on antenna timing measurements is a Gaussian with covariance matrix $\rbm Q$. The vector of timings is noted $\rbm T$ and the positions are noted $\rbm P$. This let us write the matrix $\rbm \Sigma = (\rbm P^T \rbm Q^{-1} \rbm P)^{-1}$. Our objective is to determine :
\begin{equation}
        \rm p \left(\ \rbm k \  |\,\rbm P, \rbm T, \, \Vert \rbm k \Vert=1  \right),
\end{equation}
which has no analytical solution. However, as we show in Sect.~\ref{section:performances}, the reconstruction errors of the estimators proposed in the previous section are small (typically less than a degree). We can assume that the unit sphere is locally flat and approximated by its tangent plane at $\rbm{k_{MLE}}$ :$\mathcal{T}_{\rbm{MLE}}$. The distribution of $\rbm k$ on the unit sphere is thus locally approximated by the distribution on $\mathcal{T}_{\rbm{MLE}}$.

    \begin{equation}
        \rm p \left(\ \rbm k \  |\,\rbm P, \rbm T, \, \Vert \rbm k \Vert=1  \right) \approx 
        \rm p \left(\ \rbm k \  |\,\rbm P, \rbm T, \, \rbm k\in\mathcal{T}_{\rbm{MLE}}\right).
    \end{equation}
    
    We will write the previous vectors in the basis $\rbm{b}=(\rbm{\hat{e_r}}, \rbm{\hat{e_\theta}}, \sin \rm \theta\ \rbm{\hat{e_\phi}})$, the spherical basis associated with $\rbm{k_{MLE}}$ scaled by $\sin \theta$ along $\rbm{\hat{e_\phi}}$. We will add a subscript $\rm sph$ to vectors expressed in this basis.
    
    Let us denote by $\rbm R$ the transition matrix: 
    \begin{equation}
        \rbm{k_{sph ,MLE}} = \rbm{R}  \, \rbm{k_{MLE}} = 
            \left(\begin{matrix}
                1 \\ 0 \\ 0
            \end{matrix}\right)\ .
    \end{equation}

    The basis $\rbm b$ is such that the tangent plane $\mathcal{T}_{\rbm{MLE}} = \left\{ \rbm{k_{sph}} \in \mathbb{R}^3\rbm |\ \rbm{k_{sph}}\cdot\rbm{\hat{e_r}} = 1 \right\}$ and that
    $(\rbm{\hat{e_\theta}}, \sin \rm \theta\ \rbm{\hat{e_\phi}})$ is a basis of $\mathcal{T}_{\rbm{MLE}}$.
    Furthermore, for small variation $\rbm{dk_{sph}}$ close to $\rbm{k_{sph ,MLE}}$ on $\mathcal{T}_{\rbm{MLE}}$ we have:
    \begin{equation}
         \rbm{dk_{sph}} = \left(\begin{matrix}
                 0 \\ \rm d \theta \\ \rm d \phi
             \end{matrix}\right) + \, o( \rbm{dk_{sph}})
             = \left(\begin{matrix}
                 0 \\ \rbm{k_a}
             \end{matrix}\right) + \, o( \rbm{dk_{sph}}),
     \end{equation}
    where $\rm d \theta$ and $\rm d \phi$  are small variations of zenith and azimuth angles.

    Rewriting \ref{eq:DistNoConstraint} in this new basis gives:
    \begin{equation}
        \rbm{k_{sph}} 
        \thicksim \mathcal{N}\left( \rbm {k_{sph}^*},\ \rbm R \rbm \Sigma \rbm R^T \right)\ .
    \end{equation}
        
    Let $\rbm{k_a} \in \mathbb{R}^{2\times1}$ and $k_r \in \mathbb{R}$
    such that $\rbm{k_{sph}} = [k_r,\,\rbm{k_a}^T]^T$. Let $\rbm{k_a^*} \in \mathbb{R}^{2\times1}$ and $k_r^* \in \mathbb{R}$ such that $\rbm{k_{sph}^*} =[k_r^*,\,{\rbm{k_a^*}}^T]^T$. Finally, let $\rbm{\Sigma_{rr}} \in\mathbb{R}$, $\rbm{\Sigma_{aa}} \in \mathbb{R}^{2\times2}$ and $\rbm{\Sigma_{ar}} \in \mathbb{R}^{2\times1}$.
    such that:
    \begin{equation}
    \rbm{R} \rbm{\Sigma} \rbm{R}^T = \left(
    \begin{matrix}
        \rbm{\Sigma_{rr}} & \rbm{\Sigma_{ar}}^T \\
        \rbm{\Sigma_{ar}} & \rbm{\Sigma_{aa}} 
    \end{matrix} \right)\ .
    \end{equation}
    Each of these quantities are denoted by $\rbm r$ if they involve the coordinate along $\rbm{\hat{e_r}}$ and by $\rbm a$ if they involve a coordinate corresponding to an angle (along  $\rbm{\hat{e_\theta}}$ or $\sin \rm \theta\ \rbm{\hat{e_\phi}}$). For example, $\rbm{\Sigma_{rr}}$ is the variance of the distribution along $\rbm{\hat{e_r}}$. 
    
    The condition "$\rbm k\in\mathcal{T}_{\rbm{MLE}}$" translates into "$k_{\rm r}=1$". The conditional distribution is Gaussian with covariance matrix: 
    \begin{eqnarray}\label{eq:Scur}
            \Bar{\bm \Sigma} &=& \rm{Cov}(\rbm{k_a} \, \vert \, k_{\rm r}=1 ) = \rm{Cov}\left(\left(\begin{matrix}
                \theta \\  \phi
            \end{matrix}\right)\right) \nonumber\\
            &=& \rbm{\Sigma_{aa}} - \frac{1}{\rbm{\Sigma_{rr}}}\rbm{\Sigma_{ar}}^T\rbm{\Sigma_{ar}}
    \end{eqnarray}

    To summarize, the final distribution of the reconstructed zenith and azimuth is given by:
    
    \begin{equation}
        \left(\begin{matrix} \theta \\ \phi\end{matrix}\right)  \thicksim 
        \mathcal{N}\left( \left(\begin{matrix} \theta_{\rm MLE} \\ \phi_{\rm MLE} \end{matrix}\right),\, \Bar{\rbm \Sigma} \right),
    \end{equation}
where $\theta_{\rm MLE}$ and $\phi_{\rm MLE}$ are the zenith and azimuth angles of $\rbm{k_{MLE}}$.

   The expression for $\Bar{\rbm \Sigma}$ in Eq.~\ref{eq:Scur} is the Schur complement of the block matrix $\rbm{R} \rbm{\Sigma} \rbm{R}^T$. Its inverse is the corresponding block in the block matrix $\left(\rbm{R} \rbm{\Sigma} \rbm{R}^T\right)^{-1}$. We thus have:
    \begin{equation}
        \Bar{\rbm \Sigma} = 
        \left[  
            \rbm{R_a}^T \rbm \Sigma^{-1} \rbm{R_a}
        \right]^{-1}
    \end{equation}
        with
        \begin{equation}
        \rbm{R_a} = \left(\begin{matrix}
            -\cos(\theta)\cos(\phi)& \sin(\theta)\sin(\phi) \\
            -\cos(\theta)\sin(\phi)& -\sin(\theta)\cos(\phi) \\
                       \sin(\theta)   &              0        \\
        \end{matrix} \right).
    \end{equation}

    In practice, we approximate $\rbm{R_a}$ by replacing $\theta$ by $\theta_{\rm{MLE}}$ and $\phi$ by $\phi_{\rm{MLE}}$, the zenith and azimuth angles of $\rbm{k_{MLE}}$.

\section{Refractive index refinement}\label{app:IRrefinement}
To model the wavefront, a proper description of the wave propagation speed is necessary. This speed depends on the medium traversed by the signal. If the EAS starts to develop high in the atmosphere, the mean refractive index would be smaller than for a EAS that starts closer to the ground where the atmosphere density is larger. Given an event, the mean refractive index is different for every antenna.

\begin{figure}[ht]
    \centering
    \includegraphics[scale=0.5]{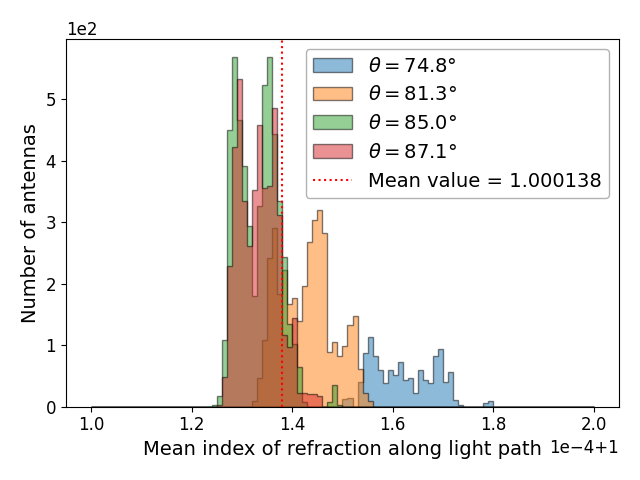}
    \caption{Distributions of mean refractive indices for all events and all antennas in the ZHAireS simulation set. The refractive index calculated is the mean index between $X_{\rm max}$ position, a proxy of the virtual source position, and the antenna position.}
    \label{fig:Idxofrefract}
\end{figure}

\begin{figure*}[ht]
    \centering
    \includegraphics[scale=0.49]{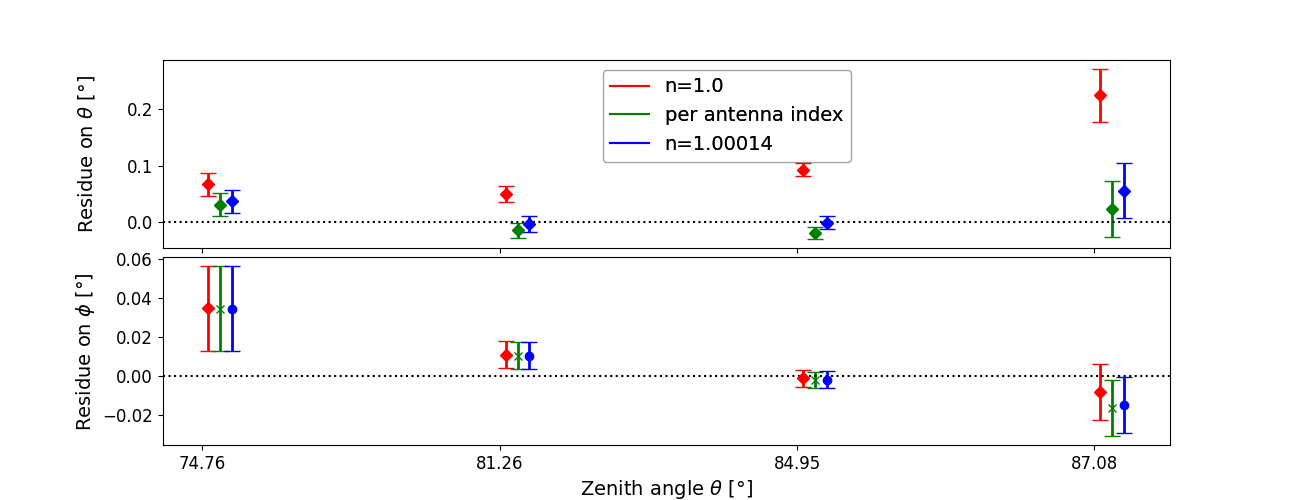}
    \caption{Mean zenith (top) and azimuth (bottom) angle reconstruction residues on ZHAireS simulation set using a refractive index $n_{\rm{atm}}=1$ (red), $n_{\rm{atm}}=1.00014$ (blue), and a refractive index varying for every antenna (green).}
    \label{fig:perfPWFZHAireSIRRefinement}
\end{figure*}
To compute this refractive index along the light path, we consider that the signal travels in a straight line and was emitted at the location of $X_{\rm max}$ \cite{Decoene_2023}. 
The effective refractive index model used is the same as the one in the ZHAireS simulations \cite{Alvarez_Mu_iz_2012}.
We represent on Fig.~\ref{fig:Idxofrefract} the histograms of the mean refractive indices along the light path for every event and every antenna in the ZHAireS simulation set. 

We present in Fig.~\ref{fig:perfPWFZHAireSIRRefinement} the binned residues for $\theta$ and $\phi$ reconstruction o the ZHAireS simulation set using the semi-analytical method, where we consider a refractive index $n_{\rm{atm}}=1$ (red), $n_{\rm{atm}}=1.00014$ (blue), and a refractive index varying for every antenna (green).
As can be seen, it is necessary to account for the atmosphere density. However using an average value for all the events and antennas provides satisfactory results, and alleviates the need to recompute the refractive index for every antenna.

\end{document}